\documentclass{article}

\usepackage{exscale}
\usepackage{latexsym}
\usepackage{amsmath}
\usepackage{amssymb}
\usepackage[OT1]{fontenc}
\usepackage[latin1]{inputenc}
\usepackage{graphics}
\usepackage{graphicx}
\usepackage{color}

\begin{document}

\markboth{Lichtenegger, Schappacher}
{A Carbon-Cycle Based Stochastic CA Climate Model}

\title{A CARBON-CYCLE BASED STOCHASTIC
 CELLULAR AUTOMATA CLIMATE MODEL}
 
\date{March $23^{\mathrm{rd}}$, 2011}

\author{KLAUS LICHTENEGGER,\\
\small Institut f\"ur Physik, Karl-Franzens-Universit\"at Graz,\\
\small Institut f\"ur Analysis und Computational Number Theory,\\
\small Technische Universit\"at Graz;
\small A-8010 Graz, Austria,\\
\small klaus.lichtenegger@uni-graz.at\\[12pt]
WILHELM SCHAPPACHER,\\
\small Institute of Mathematics and Scientific Computing, Heinrichstra\ss{}e 36 \\
\small Graz University, 8010 Graz, Austria\\
\small wilhelm.schappacher@uni-graz.at}

\maketitle

\begin{abstract}
\noindent In this article a stochastic cellular automata model
is examined, which has been developed to study a ``small''
world, where local changes may noticeably alter global
characteristics. This is applied to a climate model, where
global temperature is determined by an interplay between
atmospheric carbon dioxide and carbon stored by plant life.
The latter can be relased by forest fires, giving rise to
significant changes of global conditions within short time.

\end{abstract}

\section{Introduction} 
\label{klsec:carbclim_intro}
%%%%%%%%%%%%%

\subsection{A Brief Note on Cellular Automata}

Cellular Automata (CA), as studied first by von Neumann and
Ulam~\cite{vNThSelfReprod}, provide systems both interesting
from a fundamental point of view and for practical purposes of model
builders. Among the systems most frequently studied are CA versions
of daisyworld~\cite{Watson:1983daisy, Gingerbooth:Sim, CataDaisy} and forest fire
models~\cite{BCT:1990, CBJ:1990, clar-1994, clar-1996-8, KreHerg09}.

The main idea is to study a grid or network of identical units
(cells), which each can take discrete states, and where the time evolution
of each cell is governed by simple local rules: The state $s_i(t)$ of a
cell labelled by $i$ at timestep $t$ is only determined by its previous
state $s_i(t-1)$ and the states $s_{j_k}(t-1)$ of a small
number $k$ of neighbours, labelled by $j_k$.

There are different ways how to precisely define a neighbourhood,
but typically the qualitative behaviour is typically unaffected by the
details~\cite{Lichtenegger:2005DA, LiSchapp:2009Forest}.
This can be seen as a footprint of \emph{universality} -- the fact that
microscopic details do not influence features of systems which are
``critical'' in the sense of Wilson~\cite{Wilson:1973jj}.
The fact that all rules are local implies a finite maximum speed
for the propagation of all effects.

Originally the rules governing cellular automata were
strictly deterministic. One can extend this formalism (and make
it both more flexible and more realistic) when including stochastic
elements, i.e. give \emph{probabilities} how the state of a cell
depends on its own previous state and those of its
neighbours. On can also replace the few discrete states by a
(quasi)\-continuum to gain additional flexibility. The system studied
in this article belongs to this class of Stochastic Cellular
Automata (SCA) models, which the authors apply to a particular
feature of climate-relevant dynamics.

\subsection{A Brief Note on the Carbon Cycle}
\label{klssec:intro_carbcyc}

It is well-known that the global temperature depends strongly on
the amount of greenhouse gases contained in the atmosphere.
While $\mathrm{H}_2\mathrm{O}$ vapor is the most important of them,
its dynamic is extremely complicated and hard to treat reliably
in models. Since the amount of $\mathrm{H}_2\mathrm{O}$ vapor
strongly depends on temperature, the water vapor cycle may
largely act as a positive feedback cycle, enhancing the effects
of other greenhouse gases.

For these reasons the authors focus on $\mathrm{CO}_2$, being
the most important other greenhouse gas. Atmospheric
$\mathrm{CO}_2$ is crucial for plant life and one typically expects
enhanced growth for larger amount of $\mathrm{CO}_2$ available.
This is indeed sometimes used as an argument against global
warming~\cite{Fowler:PosEffect}.

But in particular when accompanied by climate changes,
enhanced plant growth can also mean increased vulnerability
to large-scale forest fire, which can release considerable
amounts of $\mathrm{CO}_2$ in very short time~\cite{FireClimate}
(though balancing aspects seem possible~\cite{FireCooling}).

There exists a number of realistic climate models, which attempt
to directly describe the terrestric system~\cite{Clim07Models}.
These models, which are by now quite successful, are mostly
based on the (simplified) treatment of atmosphere and ocean
dynamics, but mostly neglect the plant life Carbon cyle.
Therefore the present model, though significantly more simplistic,
can be regarded as complementary to these approaches.

%\newpage

%%%%%%%%%%%%%%%%%%%%%%%%%%%%%%%%%%%
\section{The Model}
\label{klsec:carbclim_model}

\subsection{The Basic Idea}

In this model one studies patches of land, which are, in the course of time,
inhabited by plant life. During their growth process, these plants can store
carbon from the atmosphere. The growth rate depends both on temperature
(determined by geographical latitude and carbon dioxide in the atmosphere)
and the amount of carbon available (i.e. again carbon dioxide in the atmosphere).

For each cell, optimum growth temperature and temperature tolerance are
first set randomly and then subject to a simple evolutionary process.
Uninhabited cells can become inhabited by spreading from neighbours.
Their characteristic parameters are determined as a weighted average of
those of all neighbours, affected by mutations.

In addition, there are spontaneous outbreaks of forest fires, and patches are
more vulnerable to them, if they are more heavily overgrown. The fires can
spread, and burning cells immediately release all carbon stored in them to
the atmosphere in form of carbon dioxide.

\smallskip

\noindent \textbf{Remark}: Since the amount of oxygen in the atmosphere is
not noticably changed during all processes, in the following the distinction
between ``carbon'' and ``carbon dioxide'' will be dismissed and always the
word ``carbon'' is used, where it is implicitly assumed that a sufficient amount
of oxygen is available for all processes and that carbon is deposited in the
atmosphere in the form of carbon dioxide.

In the process of biological degradation and during fires, not only
$\mathrm{CO}_2$, but also considerable amounts of $\mathrm{H}_2\mathrm{O}$
are released. While, for reasons discussed in section~\ref{klssec:intro_carbcyc}
we do not attempt to explicitly include water vapor effects, they
may be -- at least at some timescales -- partially covered too.

\subsection{Relation to Previous Work}

The system studied in this approch is closely related to
the Clar-Drossel-Schwabl forest-fire
model~\cite{clar-1994, clar-1996-8} which
also combines slow growth with rapid burning of
clusters. In contrast to the Clar-Drossel-Schwabl model,
the present approach features stochastic propagation of
the fire and -- more important -- the coupling to a global
variable, to be interpreted as the atmospheric carbon
dioxide content of the model world.

\subsection{Implementation}
\label{ssec:implementation}

The model is set up on a rectangular grid, described by several
$[N_i\times N_j]$-matrices, which are summarized and explained
in table~\ref{kltab:CCCmod_sysmat}. Each cell represents a patch
of land, possibly inhabited by a certain amount of plant life with
individual characteristics. Boundary conditions are chosen periodic
in the first ($i$) and open in the second ($j$) direction.

\begin{figure}
  \begin{center}
  \begin{tabular}{|c|l|l|} \hline
    matrix & range & meaning \\ \hline
    $\mathbf{L}$ & $\ell_{ij}\in\{0,\,1\}$ & life/inhabitance Matrix \\[3pt]
    $\mathbf{C}$ & $c_{ij}\in[0,\,1]$ & amount of carbon stored in a specific cell \\[3pt]
    $\mathbf{T}$ & $\tau_{ij}\in \mathbb{R}$ & optimal temperature for a cell \\[3pt]
    $\mathbf{S}$ & $\sigma_{ij}\in \mathbb{R}^+$ & temperature tolerance
       (width of distribution) \\[3pt]
    $\mathbf{B}$ & $b_{ij}\in\{0,\,1\}$ & flag for burning cell (used only in
       forest fire phase) \\ \hline
  \end{tabular}
  \end{center}
  \caption{System matrices}
  \label{kltab:CCCmod_sysmat}
\end{figure}

\bigskip

\noindent A simulation run is performed as follows:

\smallskip

\textbf{Initialization}: Cells are inhabited with starting density $\rho_{\mathrm{start}}$,
the parameters of inhabited cells are determined as $c_{ij}=U$,
$\tau_{ij}=U$ and $\sigma_{ij}=1+\mu\cdot (2U-1)$ with $U$ taken
individually from a uniform $[0,\,1)$-distribution. Now the following cycle
is repeated $N_{\mathrm{steps}}$ times:

\begin{enumerate}
  \item \textbf{Spreading}: Each cell with at least one inhabited neighbour
    has a chance to become inhabited as well. The probability is given
    by\footnote{In principle, one could introduce an additional multiplicative
    parameter $\alpha_{\mathrm{spread}}\in(0,\,1]$ in front of the sum, the same
    way it is done for the following processes. Our choice of
    $\alpha_{\mathrm{spread}}=1$ effectively fixes the timescale. Accordingly
    the rates of all other processes (except spreading of forest fires) are determined
    relative to this spreading rate.}
    \begin{equation}
      p(i,\,j) = \frac1{4+4\pi_\mathrm{M}}\sum_{i'j'}{}' \,c_{i'j'}\,.
    \end{equation}
    where the prime at the sum indicates summation over next neighbours,
    with diagonal elements ($i\pm1,\,j\pm1$) weighted with the Moore
    factor~\cite{Lichtenegger:2005DA} $\pi_\mathrm{M}=\frac12$. In case the cell
    becomes inhabited, its parameters are determined as
    \begin{align}
      \tau_{ij} &= \frac1N\,\sum_{i'j'}{}' \,c_{i'j'}\tau_{i'j'} +\mu\cdot(2U-1)\,, \\
      \sigma_{ij} &= \frac1N\,\sum_{i'j'}{}' \,c_{i'j'}\tau_{i'j'}\,\left( 1+\mu\cdot(2U-1)\right)\,,
    \end{align}
    with $N:=\sum{}' _{i'j'}\,c_{i'j'}$ and $U$ individually taken from a uniform
    random distribution on $[0,\,1)$. The row-averaged values
    \begin{equation}
      \tau_{j} = \frac1{\sum_i \ell_{ij}}\, \sum_{i=1}^{N_i} \ell_{ij}\tau_{ij}
      \qquad\mathrm{and}\qquad
      \sigma_{j} = \frac1{\sum_i \ell_{ij}}\, \sum_{i=1}^{N_i} \ell_{ij}\sigma_{ij}
    \end{equation}
    are stored for later time series analysis.
    
  \item \textbf{Determining global variables (1)}:
    The carbon content $C$ of the atmosphere is determined as
    \begin{equation}
      C = 1 - \frac1{N_i\,N_j}\,\sum_{i,j} c_{ij}
      \label{kle:CCCmod_getC}
    \end{equation}
    and the temperature is determined as
    \begin{equation}
      T_j = \alpha_{\mathrm{Tvar}}\,\frac{j-1}{N_j-1} + (1-\alpha_{\mathrm{Tvar}})\,C\,.
      \label{kle:CCCmod_getT}
    \end{equation}
    The value of $C$ is stored for later time-series analysis.
  \item \textbf{Growth}: Each cell growths with its individual rate,
    \begin{equation}
      c_{ij}(t+1) = \mathrm{med}\Big\{ 0,\, c_{ij}(t)+\alpha_{\mathrm{growth}}\,C\,
         \varphi(T_j,\,\tau_{ij},\,p_{ij}),\, 1 \Big\} 
    \end{equation}
    (where ``med'' denotes the median value) with the temperature
    adaptation model function
    \begin{equation}
      \varphi(T_j,\,\tau_{ij},\,\sigma_{ij})
         = N \exp\left\{ - \frac{(T_j-\tau_{ij})^2}{2\sigma_{ij}}\right\} - h_{\mathrm{hostil}}\,,
      \label{kle:carbmod_gaussian}
    \end{equation}
    where the normalization $N=N(\tau_{ij},\,\sigma_{ij})$ is chosen such that
    \begin{equation}
       \int_0^1\varphi(T_j,\,\tau_{ij},\,\sigma_{ij}) = 1 - h_{\mathrm{hostil}}\,.
    \end{equation}
    This simple Gaussian shape already shows the most important features
    one would expect of such a function. It allows both extreme specialization
    to a certain temperature and adaptation to a wider range.
    For $h_{\mathrm{hostil}}>0$ there may be temperature regions where
    growth is negative. If this causes $c_{ij}(t+1)$ to be $= 0$, the cell dies.
    
  \item \textbf{Determining global variables (2)}:
    Again Carbon content and temperature are determined according to
    equations~\eqref{kle:CCCmod_getC} and~\eqref{kle:CCCmod_getT}.

  \item \textbf{Forest fires}:
    Each inhabited cell $(i,\,j)$ has a chance to catch fire, with probability
    \begin{equation}
      p = \alpha_{\mathrm{fire}}\,T_j\,c_{ij}\,.
    \end{equation}
    If there any burning cells (cells that caught fire), a complementary
    simulation on a short timescale is started. The fire may spread from
    burning cells ($b_{ij}=1$) to their neighbours: 
    \begin{enumerate}
      \item In each step of the burning process, each inhabited cell can
      catch fire with probability
      \begin{equation}
        p = T_j\,c_{ij}\,\frac1{4+4\pi_\mathrm{M}}\sum_{i'j'}{}' \,b_{i'j'}\,.
      \end{equation}
      \item In each individual cell, the fire lasts for one timestep and is
      extinguished afterwards ($b_{ij}\to 0$). A burning cell which has
      been burning already in the last step is set back to uninhabited
      state.
    \end{enumerate}
    
\end{enumerate}

The system parameters are summarized in figure~\ref{kltab:carbclim_syspar}.

\begin{figure}
 \begin{center}
  \begin{tabular}{|c|l|l|} \hline
    param. & range & meaning \\ \hline
    $N_i$ & $\in\mathbb{N}_{\ge 2}$ & linear size of the grid in first direction (open b.c.) \\[6pt]
    $N_j$ & $\in\mathbb{N}_{\ge 2}$ & linear size of the grid in second direction (periodic b.c.)  \\[6pt]
    $t_{\mathrm{max}}$ & $\in 2^{\mathbb{N}}$ & simulation time \\[6pt]
    $\rho_{\mathrm{start}}$ & $\in(0,\,1]$ & initial density of inhabited cells \\[6pt]
    $\alpha_{\mathrm{fire}}$ & $\in\mathbb{R}^+$ & %measure for
       probability of a cell spontaneously catching fire \\[6pt]
    $\alpha_{\mathrm{growth}}$ & $\in\mathbb{R}^+$ & measure for growth rate \\[6pt]
    $h_{\mathrm{hostil}}$ & $\in\mathbb{R}_{\ge 0}$ & ``hostility'' of environment \\[6pt]
    $\alpha_{\mathrm{Tvar}}$ & $\in[0,\,1]$ & Geographic variability of the temperature~\eqref{kle:CCCmod_getT} \\[6pt]
    $\mu$ & $\in[0,\,1]$ & global mutation rate  \\ \hline  
  \end{tabular}
  \end{center}
  \caption{A summary of the system parameters, which are explained
    in detail in sec.~\ref{ssec:implementation}}
  \label{kltab:carbclim_syspar}
\end{figure}

%%%%%%%%%%%%%%%%%%%%%%%%%%%%%%%%%%%%%
\section{Simulation and Results}

\subsection{Simulation and Data Analysis}
\label{klssec:carbclim_analysis}

\begin{figure}
  \includegraphics[width=12.7cm]{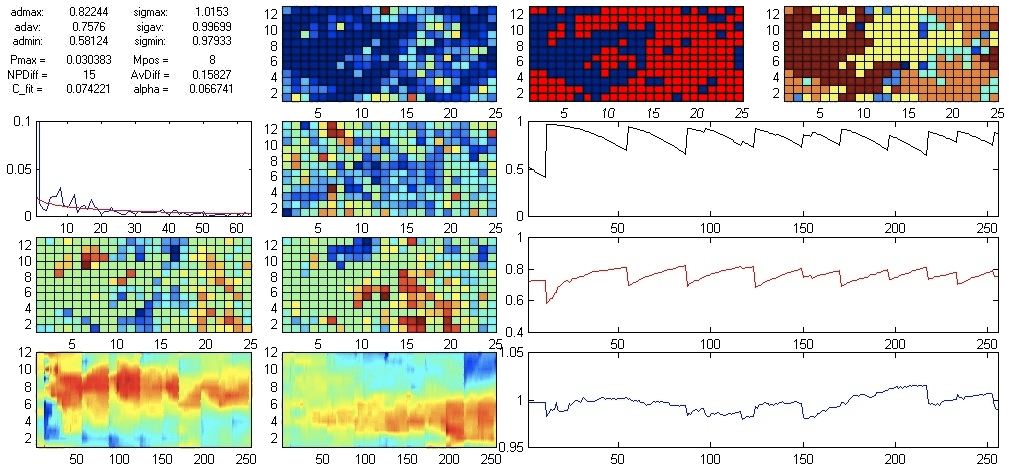}
  \caption{Results of a simulation run for the parameter choice
   $N_i=12$, $N_j=24$, $t_{\mathrm{max}}=256$, $\rho_{\mathrm{start}}=0.9$,
   $\alpha_{\mathrm{fire}}=0.002$, $\alpha_{\mathrm{growth}}=0.1$,
   $h_{\mathrm{hostil}}=1$, $\alpha_{\mathrm{Tvar}}=0.5$, $\mu=0.1$.\newline
   The graphical output contains the following elements: \newline
   In the first row the system parameters are summarized, the first plot
   displays the currently stored amount of carbon, the second one
   shows currently inhabited (red/light) and uninhabited (blue/dark) cells and the
   third one is a graphical representation of the last forest fires. \newline
   In the second row one has the Fourier spectrum (absolute values),
   including a power-law fit according to equation~\eqref{eq:powerlawfit},
   an illustration of the total amount of carbon stored in the cells
   over the whole simulation run and the time series of carbon in
   the atmosphere. \newline
   The third row contains the current $\tau$-values, the current
   $\sigma$-values and a time series of average temperature
   adaptation~\eqref{eq:adapt}.\newline
   The fourth row displays the time series of
   row-averaged $\tau$ and $\sigma$ and a time series of
   average versatility~\eqref{eq:versat}. (In case that row-averaged values
   were not well-defined for certain times [when no cell of that particular row
   was inhabited], this is reflected by white spots in the corresponding graphs,
   cf. figure.~\ref{klfig:mutation}.)
   }\label{klfig:simrun}
\end{figure}

The result of a typical simulation run is displayed in figure~\ref{klfig:simrun}.
After a simulation run, the following data is available:
\begin{itemize}
\item The time series of carbon $C$ in the atmosphere,
\item the time series of average temperature adaptation
  \begin{equation}
    A(t) := 1-\frac1{N_L(t)}\sum_{ij} \ell_{ij}(t)\,\left| \tau_{ij}(t)-T_j(t) \right|\,,
    \label{eq:adapt}
  \end{equation}
\item the time series of average versatility
  \begin{equation}
    V(t) := \frac1{N_L(t)}\sum_{ij} \ell_{ij}(t)\sigma_{ij}(t)\,,
    \label{eq:versat}
  \end{equation}
\end{itemize}
where the defintion $N_L(t) := \sum_{ij} \ell_{ij}(t)$ has been used.
From these time series, the authors have extracted maximum, average
and minimum values for $A$ and $V$ and certain characteristics
of the Fourier spectrum of $C$ (examining only absolute values,
i.e. the amplitude spectrum $X(n)$, which is closely related to the
power spectrum).

These characteristics are the position and the value of the dominant
maximum beyond $n=1$ and the parameters of a power-law fit
\begin{equation}
  P(n) \approx \frac{C}{n^{\alpha}} + o_{\mathrm{offset}}\,.
  \label{eq:powerlawfit}
\end{equation}

In addition the authors have analyzed the number of exceptional increases
of $C$ (usually caused by forest fires), which are defined as
timesteps $\tau$, where $C(\tau)>C(\tau-1)$, but $C(\tau-1)<C(\tau-2)$.
and the average over the changes $\Delta C(\tau):=C(\tau)-C(\tau-1)$.

% \newpage

\subsection{Results}
\label{klssec:carbclim_results}

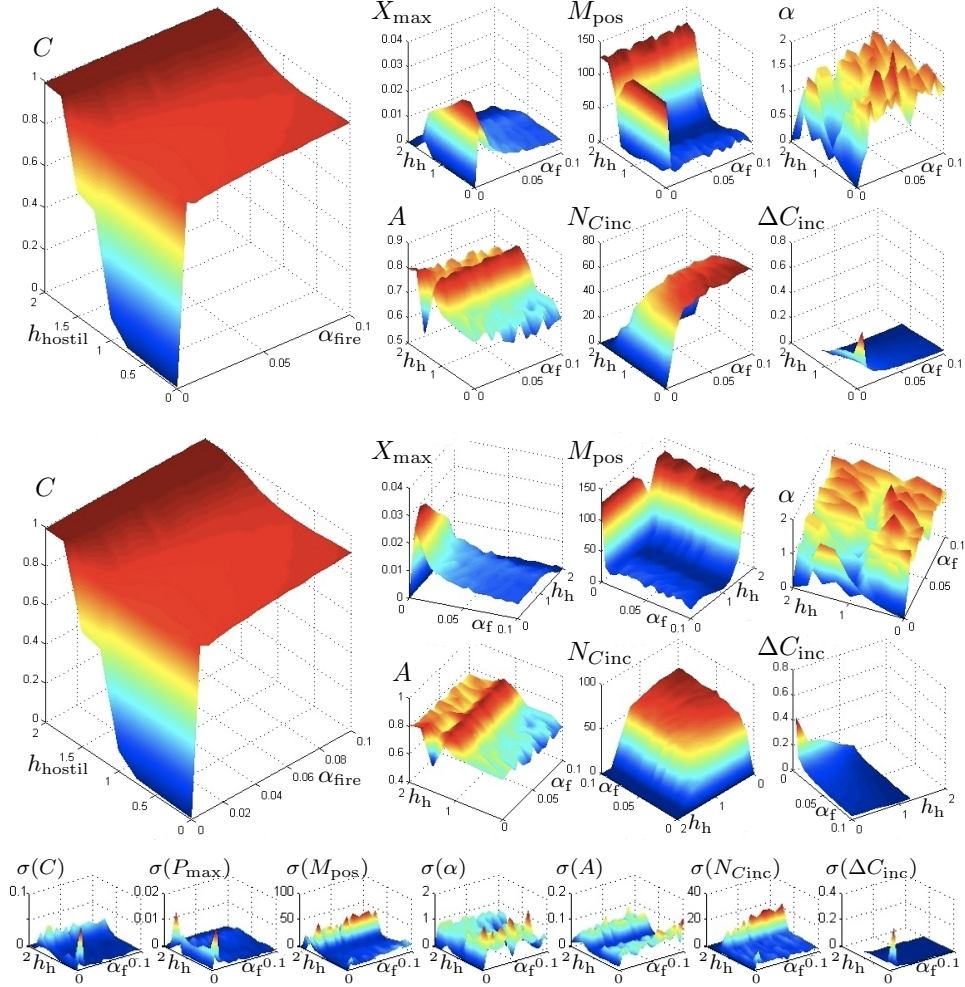
\begin{figure}
  \includegraphics[width=12.7cm]{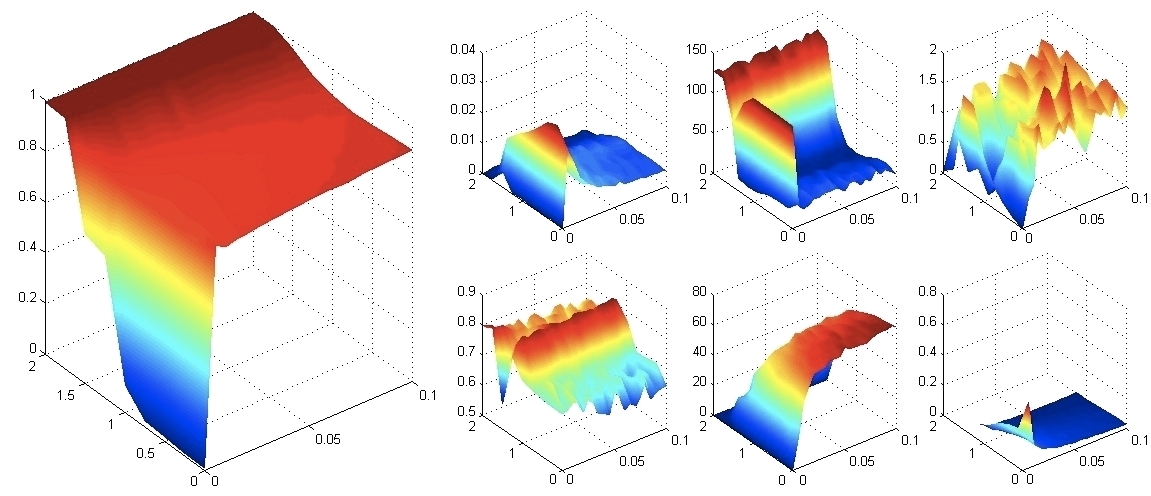}
  \put(-352,134){$C$}\put(-354,26){\small $h_{\rm hostil}$}\put(-244,26){\small $\alpha_{\rm fire}$}
  \put(-224,148){$X_{\rm max}$} \put(-150,148){$M_{\rm pos}$} \put(-70,148){$\alpha$}
  \put(-218,70){$A$} \put(-150,70){$N_{C\mathrm{inc}}$} \put(-78,70){$\Delta C_{\mathrm{inc}}$}
  %%%%%%%%%%%%%%%%%%%%%%%%%%%%%%%%%%%%%%%%%%
  \put(-212,92){\footnotesize $h_{\rm h}$}\put(-160,90){\footnotesize $\alpha_{\rm f}$}
  \put(-140,92){\footnotesize $h_{\rm h}$}\put(-88,90){\footnotesize $\alpha_{\rm f}$}
  \put(-68,92){\footnotesize $h_{\rm h}$}\put(-16,90){\footnotesize $\alpha_{\rm f}$}
  %%%%%%%%%%%%%%%%%%%%%%%%%%%%%%%%%%%%%%%%%%
  \put(-212,16){\footnotesize $h_{\rm h}$}\put(-160,14){\footnotesize $\alpha_{\rm f}$}
  \put(-140,16){\footnotesize $h_{\rm h}$}\put(-88,14){\footnotesize $\alpha_{\rm f}$}
  \put(-68,16){\footnotesize $h_{\rm h}$}\put(-16,14){\footnotesize $\alpha_{\rm f}$}\\[6pt]
  %%%%%%%%%%%%%%%%%%%%%%%%%%%%%%%%%%%%%%%%%%
  \includegraphics[width=12.7cm]{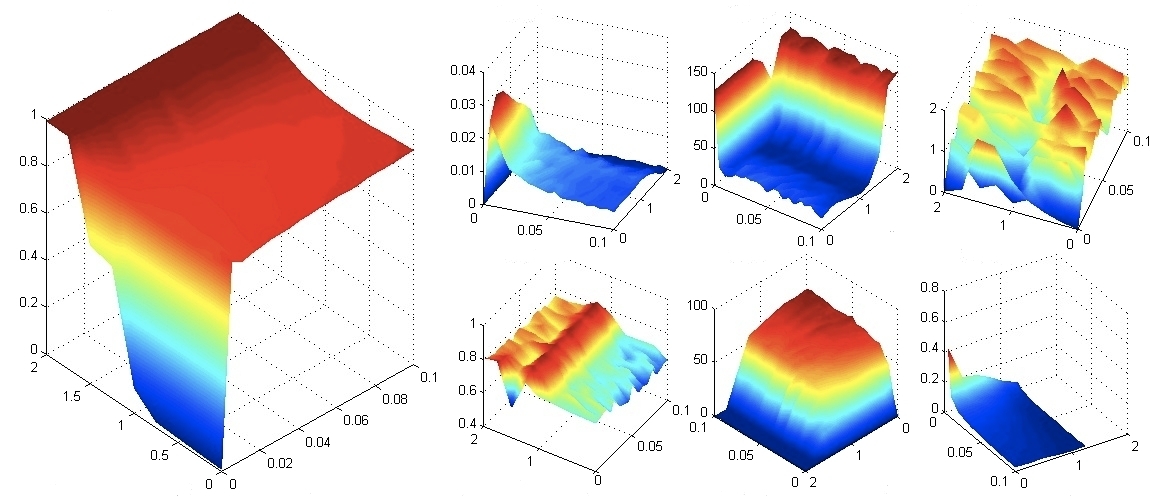}
  \put(-351,130){$C$}\put(-354,26){\small $h_{\rm hostil}$}\put(-244,22){\small $\alpha_{\rm fire}$}
  \put(-224,144){$X_{\rm max}$} \put(-150,144){$M_{\rm pos}$} \put(-70,128){$\alpha$}
  \put(-216,60){$A$} \put(-150,68){$N_{C\mathrm{inc}}$} \put(-78,70){$\Delta C_{\mathrm{inc}}$}
  %%%%%%%%%%%%%%%%%%%%%%%%%%%%%%%%%%%%%%%%%%
  \put(-156,92){\footnotesize $h_{\rm h}$}\put(-186,80){\footnotesize $\alpha_{\rm f}$}
  \put(-86,92){\footnotesize $h_{\rm h}$}\put(-120,82){\footnotesize $\alpha_{\rm f}$}
  \put(-64,86){\footnotesize $h_{\rm h}$}\put(-10,104){\footnotesize $\alpha_{\rm f}$}
  %%%%%%%%%%%%%%%%%%%%%%%%%%%%%%%%%%%%%%%%%%
  \put(-210,14){\footnotesize $h_{\rm h}$}\put(-156,20){\footnotesize $\alpha_{\rm f}$}
  \put(-104,4){\footnotesize $h_{\rm h}$}\put(-138,18){\footnotesize $\alpha_{\rm f}$}
  \put(-16,10){\footnotesize $h_{\rm h}$}\put(-56,9){\footnotesize $\alpha_{\rm f}$}\\[6pt]
  %%%%%%%%%%%%%%%%%%%%%%%%%%%%%%%%%%%%%%%%%%
  \includegraphics[width=12.7cm]{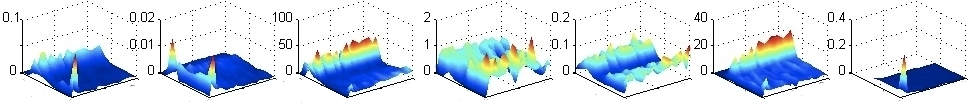}
  \put(-358,41){\footnotesize$\sigma(C)$}
  \put(-352,5){\footnotesize $h_{\rm h}$}\put(-324,4){\footnotesize $\alpha_{\rm f}$}
  \put(-356,8){\tiny$2$} \put(-336,0){\tiny$0$} \put(-316,6){\tiny$0.1$}
  \put(-308,41){\footnotesize$\sigma(P_{\rm max})$}
  \put(-301,5){\footnotesize $h_{\rm h}$}\put(-273,4){\footnotesize $\alpha_{\rm f}$}
  \put(-305,8){\tiny$2$} \put(-285,0){\tiny$0$} \put(-265,6){\tiny$0.1$}
  \put(-256,41){\footnotesize$\sigma(M_{\rm pos})$}
  \put(-250,5){\footnotesize $h_{\rm h}$}\put(-222,4){\footnotesize $\alpha_{\rm f}$}
  \put(-254,8){\tiny$2$} \put(-234,0){\tiny$0$} \put(-214,6){\tiny$0.1$}
  \put(-205,41){\footnotesize$\sigma(\alpha)$}
  \put(-198,5){\footnotesize $h_{\rm h}$}\put(-170,4){\footnotesize $\alpha_{\rm f}$}
  \put(-202,8){\tiny$2$} \put(-182,0){\tiny$0$} \put(-162,6){\tiny$0.1$}
  \put(-154,41){\footnotesize$\sigma(A)$}
  \put(-148,5){\footnotesize $h_{\rm h}$}\put(-120,4){\footnotesize $\alpha_{\rm f}$}
  \put(-152,8){\tiny$2$} \put(-132,0){\tiny$0$} \put(-112,6){\tiny$0.1$}
  \put(-103,41){\footnotesize$\sigma(N_{C\mathrm{inc}})$}
  \put(-96,5){\footnotesize $h_{\rm h}$}\put(-68,4){\footnotesize $\alpha_{\rm f}$}
  \put(-100,8){\tiny$2$} \put(-80,0){\tiny$0$} \put(-60,6){\tiny$0.1$}
  \put(-52,41){\footnotesize$\sigma(\Delta C_{\mathrm{inc}})$}
  \put(-46,5){\footnotesize $h_{\rm h}$}\put(-18,4){\footnotesize $\alpha_{\rm f}$}
  \put(-50,8){\tiny$2$} \put(-30,0){\tiny$0$} \put(-10,6){\tiny$0.1$}
  \caption{\small Results of simulation runs (10 runs for each specific
   combination of parameters) for the parameter
   choice $N_i=12$, $N_j=24$, $t_{\mathrm{max}}=256$, $\rho_{\mathrm{start}}=0.9$,
   $\alpha_{\mathrm{growth}}=0.1$, $\alpha_{\mathrm{Tvar}}=0.5$, $\mu=0.1$.
   The parameters $\alpha_{\mathrm{f}}=\alpha_{\mathrm{fire}}$ and $h_{\mathrm{h}}=h_{\mathrm{hostil}}$
   have been varied as $\alpha_{\mathrm{fire}}=0.005\,k$, $k=0,\ldots,20$ and
   $h_{\mathrm{hostil}}=0.25\,\ell$, $\ell=0,\ldots 8$.
   On the left hand side, the main plot shows the average carbon content
   $C$ of the atmosphere.
   On the right hand side, the first row of graphs gives
   the amplitude of the dominant Fourier mode, $X_{\rm max}$,
   its position $M_{\rm pos}$ and the exponent $\alpha$ of the spectral
   power-law fit~\eqref{eq:powerlawfit}. The second row shows the average
   temperature adaptation ($A$), the total number $N_{C\mathrm{inc}}$ of timesteps 
   for which $C$ increased (which corresponds to the number of significant forest fires)
   and the average change $\Delta C_{\mathrm{inc}}$ of carbon content
   during such events.
   Plots of the same quantities are given one more time (second graph on the left hand
   side, third and forth row on the right hand side), but usually shown from a different
   viewing position.
   In the last row the standard deviations $\sigma$ of the
   seven quantities under consideration are given.}
  \label{klfig:gridrunfull}
\end{figure}

Comparison of simulation runs for various sets of
parameters revealed that the most crucial parameters are
$\alpha_{\mathrm{fire}}$ and $h_{\mathrm{hostil}}$. Thus the authors
have performed simulations for values of these two parameters
chosen on a grid. For a wide mesh size and $\alpha_{\mathrm{Tvar}}=\frac12$,
the results are shown in figure~\ref{klfig:gridrunfull}.

Some of the most important features of the model are already clearly recognizable:
For $\alpha_{\mathrm{fire}}=0$ there are no forest fires, so for small $h_{\mathrm{hostil}}$
all carbon is stored in the cells. For large values of $h_{\mathrm{hostil}}$, it is still
possible that the whole population dies out due to lack of adaptation.
This happens in general (for the chosen set of parameters) for
$h_{\mathrm{hostil}}\ge1.75$, regardless of $\alpha_{\mathrm{fire}}$.

For $h_{\mathrm{hostil}}=0$ and $\alpha_{\mathrm{fire}}$ small, but nonzero,
one finds a local maximum of atmospheric carbon. This peak is a key
feature of the model and will be examined closely in the following analysis.
The basic explanation is that very small $\alpha_{\mathrm{fire}}$ implies long
periods without fires, in which large amounts of carbon are stored, such
that one forest fire can have devastating consequences and eliminate
almost all cells at once, releasing large amounts of carbon.

The authors find adaptation maxima at $h_{\mathrm{hostil}}\approx 1$ due to
the fact that for smaller values of $h_{\mathrm{hostil}}$ the punishment
for bad adaptation is less severe, while for even larger values of
$h_{\mathrm{hostil}}$ the population more frequently dies out completely
without having time to adapt.

\begin{figure}
  \includegraphics[width=12.7cm]{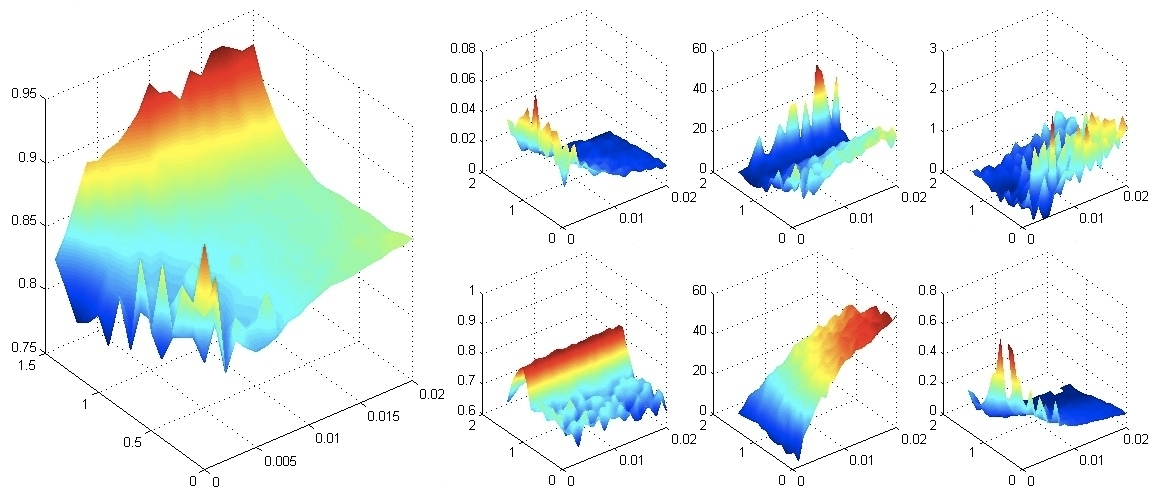}
  \put(-352,134){$C$}\put(-360,34){\small $h_{\rm hostil}$}\put(-240,28){\small $\alpha_{\rm fire}$}
  \put(-224,148){$X_{\rm max}$} \put(-150,148){$M_{\rm pos}$} \put(-70,148){$\alpha$}
  \put(-218,70){$A$} \put(-150,70){$N_{C\mathrm{inc}}$} \put(-78,70){$\Delta C_{\mathrm{inc}}$}
  %%%%%%%%%%%%%%%%%%%%%%%%%%%%%%%%%%%%%%%%%%
  \put(-212,92){\footnotesize $h_{\rm h}$}\put(-160,90){\footnotesize $\alpha_{\rm f}$}
  \put(-140,92){\footnotesize $h_{\rm h}$}\put(-88,90){\footnotesize $\alpha_{\rm f}$}
  \put(-68,92){\footnotesize $h_{\rm h}$}\put(-15,89){\footnotesize $\alpha_{\rm f}$}
  %%%%%%%%%%%%%%%%%%%%%%%%%%%%%%%%%%%%%%%%%%
  \put(-212,16){\footnotesize $h_{\rm h}$}\put(-160,14){\footnotesize $\alpha_{\rm f}$}
  \put(-140,16){\footnotesize $h_{\rm h}$}\put(-88,14){\footnotesize $\alpha_{\rm f}$}
  \put(-68,16){\footnotesize $h_{\rm h}$}\put(-15,13){\footnotesize $\alpha_{\rm f}$}\\[6pt]
  %%%%%%%%%%%%%%%%%%%%%%%%%%%%%%%%%%%%%%%%%%
  \includegraphics[width=12.7cm]{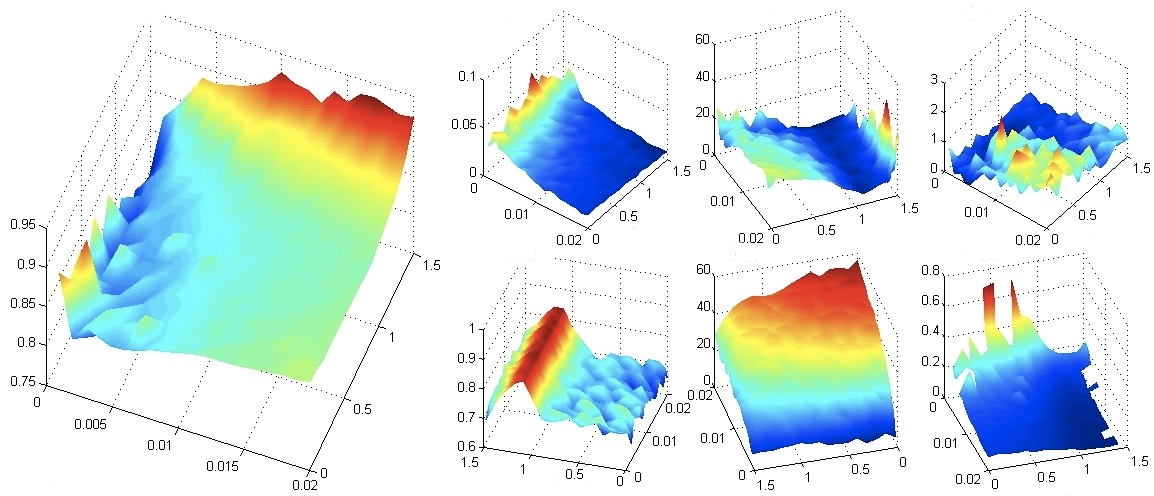}
  \put(-354,96){$C$}\put(-360,26.5){\small $h_{\rm hostil}$}\put(-234,64){\small $\alpha_{\rm fire}$}
  \put(-224,140){$X_{\rm max}$} \put(-148,148){$M_{\rm pos}$} \put(-70,136){$\alpha$}
  \put(-215,58){$A$} \put(-148,75){$N_{C\mathrm{inc}}$} \put(-78,74){$\Delta C_{\mathrm{inc}}$}
  %%%%%%%%%%%%%%%%%%%%%%%%%%%%%%%%%%%%%%%%%%
  \put(-156,96){\footnotesize $h_{\rm h}$}\put(-192,84){\footnotesize $\alpha_{\rm f}$}
  \put(-88,86){\footnotesize $h_{\rm h}$}\put(-132,86){\footnotesize $\alpha_{\rm f}$}
  \put(-50,84){\footnotesize $h_{\rm h}$}\put(-13,97){\footnotesize $\alpha_{\rm f}$}
  %%%%%%%%%%%%%%%%%%%%%%%%%%%%%%%%%%%%%%%%%%
  \put(-209,8){\footnotesize $h_{\rm h}$}\put(-156,22){\footnotesize $\alpha_{\rm f}$}
  \put(-118,2){\footnotesize $h_{\rm h}$}\put(-144,26){\footnotesize $\alpha_{\rm f}$}
  \put(-17,7){\footnotesize $h_{\rm h}$}\put(-65,11){\footnotesize $\alpha_{\rm f}$}\\[6pt]
  %%%%%%%%%%%%%%%%%%%%%%%%%%%%%%%%%%%%%%%%%%
  \includegraphics[width=12.7cm]{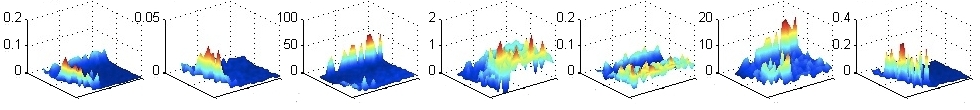}
  \put(-358,41){\footnotesize$\sigma(C)$}
  \put(-352,5){\footnotesize $h_{\rm h}$}\put(-324,4){\footnotesize $\alpha_{\rm f}$}
  \put(-356,8){\tiny$2$} \put(-336,0){\tiny$0$} \put(-316,6){\tiny$0\!.\!0\!2$}
  \put(-308,41){\footnotesize$\sigma(P_{\rm max})$}
  \put(-301,5){\footnotesize $h_{\rm h}$}\put(-273,4){\footnotesize $\alpha_{\rm f}$}
  \put(-305,8){\tiny$2$} \put(-285,0){\tiny$0$} \put(-265,6){\tiny$0\!.\!0\!2$}
  \put(-256,41){\footnotesize$\sigma(M_{\rm pos})$}
  \put(-250,5){\footnotesize $h_{\rm h}$}\put(-222,4){\footnotesize $\alpha_{\rm f}$}
  \put(-254,8){\tiny$2$} \put(-234,0){\tiny$0$} \put(-214,6){\tiny$0\!.\!0\!2$}
  \put(-205,41){\footnotesize$\sigma(\alpha)$}
  \put(-198,5){\footnotesize $h_{\rm h}$}\put(-170,4){\footnotesize $\alpha_{\rm f}$}
  \put(-202,8){\tiny$2$} \put(-182,0){\tiny$0$} \put(-162,6){\tiny$0\!.\!0\!2$}
  \put(-154,41){\footnotesize$\sigma(A)$}
  \put(-147,5){\footnotesize $h_{\rm h}$}\put(-119,4){\footnotesize $\alpha_{\rm f}$}
  \put(-151,8){\tiny$2$} \put(-131,0){\tiny$0$} \put(-111,6){\tiny$0\!.\!0\!2$}
  \put(-103,41){\footnotesize$\sigma(N_{C\mathrm{inc}})$}
  \put(-96,5){\footnotesize $h_{\rm h}$}\put(-68,4){\footnotesize $\alpha_{\rm f}$}
  \put(-100,8){\tiny$2$} \put(-80,0){\tiny$0$} \put(-60,6){\tiny$0\!.\!0\!2$}
  \put(-52,41){\footnotesize$\sigma(\Delta C_{\mathrm{inc}})$}
  \put(-46,5){\footnotesize $h_{\rm h}$}\put(-18,4){\footnotesize $\alpha_{\rm f}$}
  \put(-50,8){\tiny$2$} \put(-30,0){\tiny$0$} \put(-10,6){\tiny$0\!.\!0\!2$}
  \caption{Results of a sequence of 10 simulation runs.
  Parameters are chosen as in figure~\ref{klfig:gridrunfull},
  but only the most interesting region is examined by choosing
  $\alpha_{\mathrm{fire}}=0.001\,k$, $k=1,\ldots,20$ and
  $h_{\mathrm{hostil}}=0.1\,\ell$, $\ell=0,\ldots 15$. Again the
   graphs are displayed for two different viewing positions.\newline
   Note that for larger values of $\alpha_{\mathrm{fire}}$
   the carbon content $C$ is almost constant for small
   $h_{\mathrm{hostil}}$ and increases monotonically as
   a function of $h_{\mathrm{hostil}}$, starting at $h_{\mathrm{hostil}}\approx 1$.
   For small values of $\alpha_{\mathrm{fire}}$ and $h_{\mathrm{hostil}}$,
   there are large fluctuations in $C$, to be noticed also in the
   graph for $\sigma(C)$.}
  \label{klfig:gridrun05}
\end{figure}

\begin{figure}
  \includegraphics[width=12.7cm]{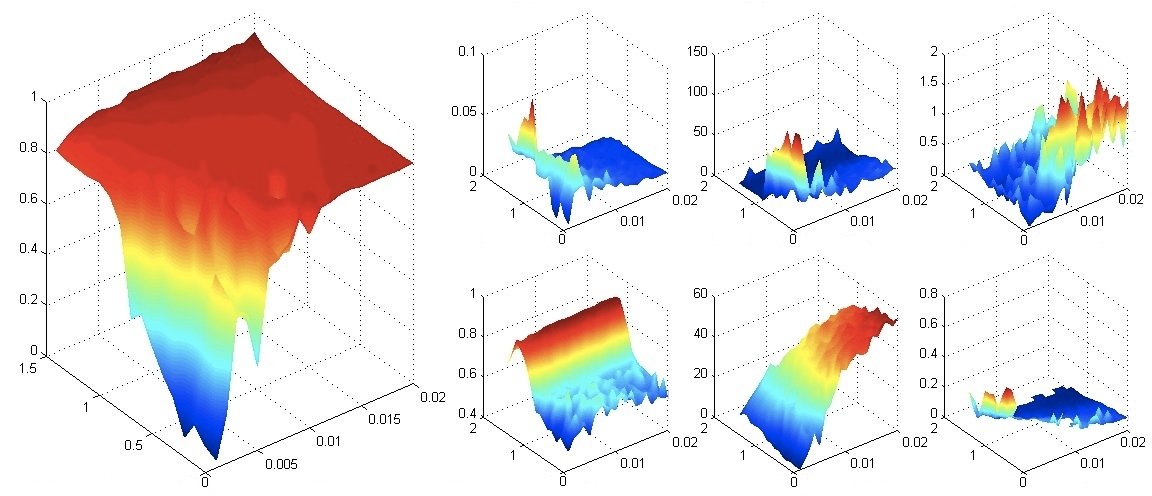}
  \put(-352,134){$C$}\put(-360,34){\small $h_{\rm hostil}$}\put(-240,27){\small $\alpha_{\rm fire}$}
  \put(-224,146){$X_{\rm max}$} \put(-150,148){$M_{\rm pos}$} \put(-70,146){$\alpha$}
  \put(-218,70){$A$} \put(-150,70){$N_{C\mathrm{inc}}$} \put(-78,70){$\Delta C_{\mathrm{inc}}$}
  %%%%%%%%%%%%%%%%%%%%%%%%%%%%%%%%%%%%%%%%%%
  \put(-212,92){\footnotesize $h_{\rm h}$}\put(-160,90){\footnotesize $\alpha_{\rm f}$}
  \put(-140,92){\footnotesize $h_{\rm h}$}\put(-88,90){\footnotesize $\alpha_{\rm f}$}
  \put(-67,92){\footnotesize $h_{\rm h}$}\put(-15,88){\footnotesize $\alpha_{\rm f}$}
  %%%%%%%%%%%%%%%%%%%%%%%%%%%%%%%%%%%%%%%%%%
  \put(-212,16){\footnotesize $h_{\rm h}$}\put(-160,14){\footnotesize $\alpha_{\rm f}$}
  \put(-140,16){\footnotesize $h_{\rm h}$}\put(-88,14){\footnotesize $\alpha_{\rm f}$}
  \put(-67,16){\footnotesize $h_{\rm h}$}\put(-15,12){\footnotesize $\alpha_{\rm f}$}\\[6pt]
  %%%%%%%%%%%%%%%%%%%%%%%%%%%%%%%%%%%%%%%%%%
  \includegraphics[width=12.7cm]{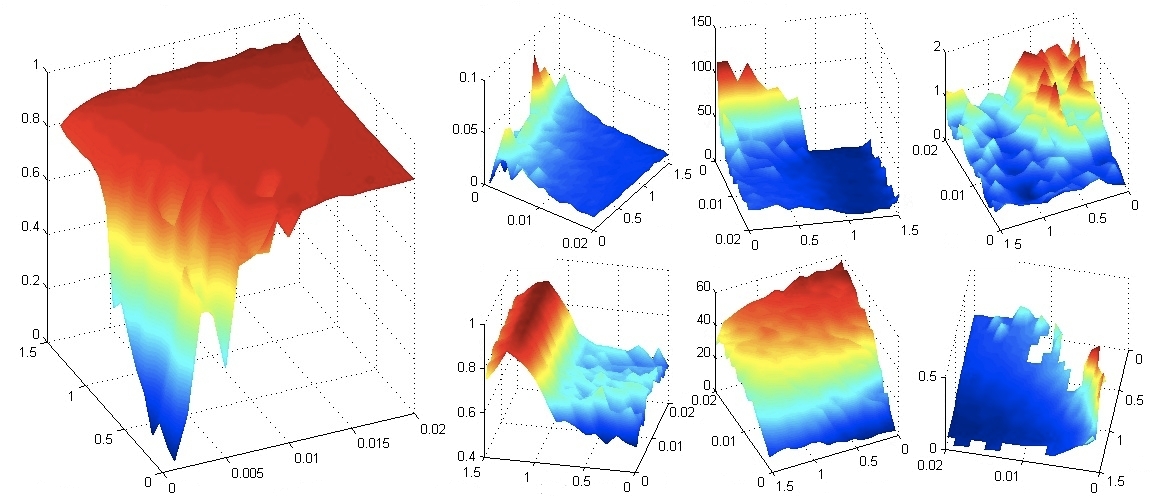}
  \put(-352,136){$C$}\put(-354,36){\small $h_{\rm hostil}$}\put(-244,10){\small $\alpha_{\rm fire}$}
  \put(-224,140){$X_{\rm max}$} \put(-138,150){$M_{\rm pos}$} \put(-68,145){$\alpha$}
  \put(-215,58){$A$} \put(-148,71){$N_{C\mathrm{inc}}$} \put(-74,74){$\Delta C_{\mathrm{inc}}$}
  %%%%%%%%%%%%%%%%%%%%%%%%%%%%%%%%%%%%%%%%%%
  \put(-155,95){\footnotesize $h_{\rm h}$}\put(-193,83){\footnotesize $\alpha_{\rm f}$}
  \put(-89,80){\footnotesize $h_{\rm h}$}\put(-140,85){\footnotesize $\alpha_{\rm f}$}
  \put(-40,78){\footnotesize $h_{\rm h}$}\put(-70,102){\footnotesize $\alpha_{\rm f}$}
  %%%%%%%%%%%%%%%%%%%%%%%%%%%%%%%%%%%%%%%%%%
  \put(-209,5){\footnotesize $h_{\rm h}$}\put(-154,20){\footnotesize $\alpha_{\rm f}$}
  \put(-112,2){\footnotesize $h_{\rm h}$}\put(-142,23){\footnotesize $\alpha_{\rm f}$}
  \put(-16,9){\footnotesize $h_{\rm h}$}\put(-64,8){\footnotesize $\alpha_{\rm f}$}\\[6pt]
  %%%%%%%%%%%%%%%%%%%%%%%%%%%%%%%%%%%%%%%%%%
  \includegraphics[width=12.7cm]{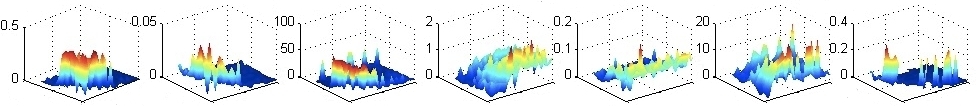}
  \put(-356,39){\footnotesize$\sigma(C)$}
  \put(-350,2){\footnotesize $h_{\rm h}$}\put(-322,1){\footnotesize $\alpha_{\rm f}$}
  \put(-354,5){\tiny$2$} \put(-333,-3){\tiny$0$} \put(-314,4){\tiny$0\!.\!0\!2$}
  \put(-308,39){\footnotesize$\sigma(P_{\rm max})$}
  \put(-299,2){\footnotesize $h_{\rm h}$}\put(-271,1){\footnotesize $\alpha_{\rm f}$}
  \put(-303,5){\tiny$2$} \put(-283,-3){\tiny$0$} \put(-263,3){\tiny$0\!.\!0\!2$}
  \put(-256,39){\footnotesize$\sigma(M_{\rm pos})$}
  \put(-248,2){\footnotesize $h_{\rm h}$}\put(-220,1){\footnotesize $\alpha_{\rm f}$}
  \put(-252,5){\tiny$2$} \put(-232,-2){\tiny$0$} \put(-212,3){\tiny$0\!.\!0\!2$}
  \put(-205,39){\footnotesize$\sigma(\alpha)$}
  \put(-196,2){\footnotesize $h_{\rm h}$}\put(-168,1){\footnotesize $\alpha_{\rm f}$}
  \put(-200,5){\tiny$2$} \put(-180,-3){\tiny$0$} \put(-160,3){\tiny$0\!.\!0\!2$}
  \put(-154,39){\footnotesize$\sigma(A)$}
  \put(-145,2){\footnotesize $h_{\rm h}$}\put(-117,1){\footnotesize $\alpha_{\rm f}$}
  \put(-149,5){\tiny$2$} \put(-129,-3){\tiny$0$} \put(-109,3){\tiny$0\!.\!0\!2$}
  \put(-103,39){\footnotesize$\sigma(N_{C\mathrm{inc}})$}
  \put(-94,2){\footnotesize $h_{\rm h}$}\put(-66,1){\footnotesize $\alpha_{\rm f}$}
  \put(-98,5){\tiny$2$} \put(-78,-3){\tiny$0$} \put(-58,3){\tiny$0\!.\!0\!2$}
  \put(-52,39){\footnotesize$\sigma(\Delta C_{\mathrm{inc}})$}
  \put(-44,2){\footnotesize $h_{\rm h}$}\put(-16,1){\footnotesize $\alpha_{\rm f}$}
  \put(-48,5){\tiny$2$} \put(-28,-3){\tiny$0$} \put(-7,3){\tiny$0\!.\!0\!2$}
  \caption{Results of a sequence of simulation runs.
   Parameters are chosen as in figure~\ref{klfig:gridrun05}
   except for $\alpha_{\mathrm{Tvar}}=0$. Again the main
   graphs are displayed for two different viewing positions.\newline
   For this choice of parameters, the temperature does not
   depend on geographical latitude, but only on $C$. Apparently
   in this situation, for small values of $\alpha_{\mathrm{fire}}$ it is more
   likely that the atmosperical carbon level remains low, i.e.
   the cells inhabitance is stable. This is also reflected in
   in the Fourier amplitude spectrum, where the maximum
   still has low intensity, indicating the absence of significant
   oscillations as they can be recognized in figure~\ref{klfig:simrun}.}
  \label{klfig:gridrun00}
\end{figure}

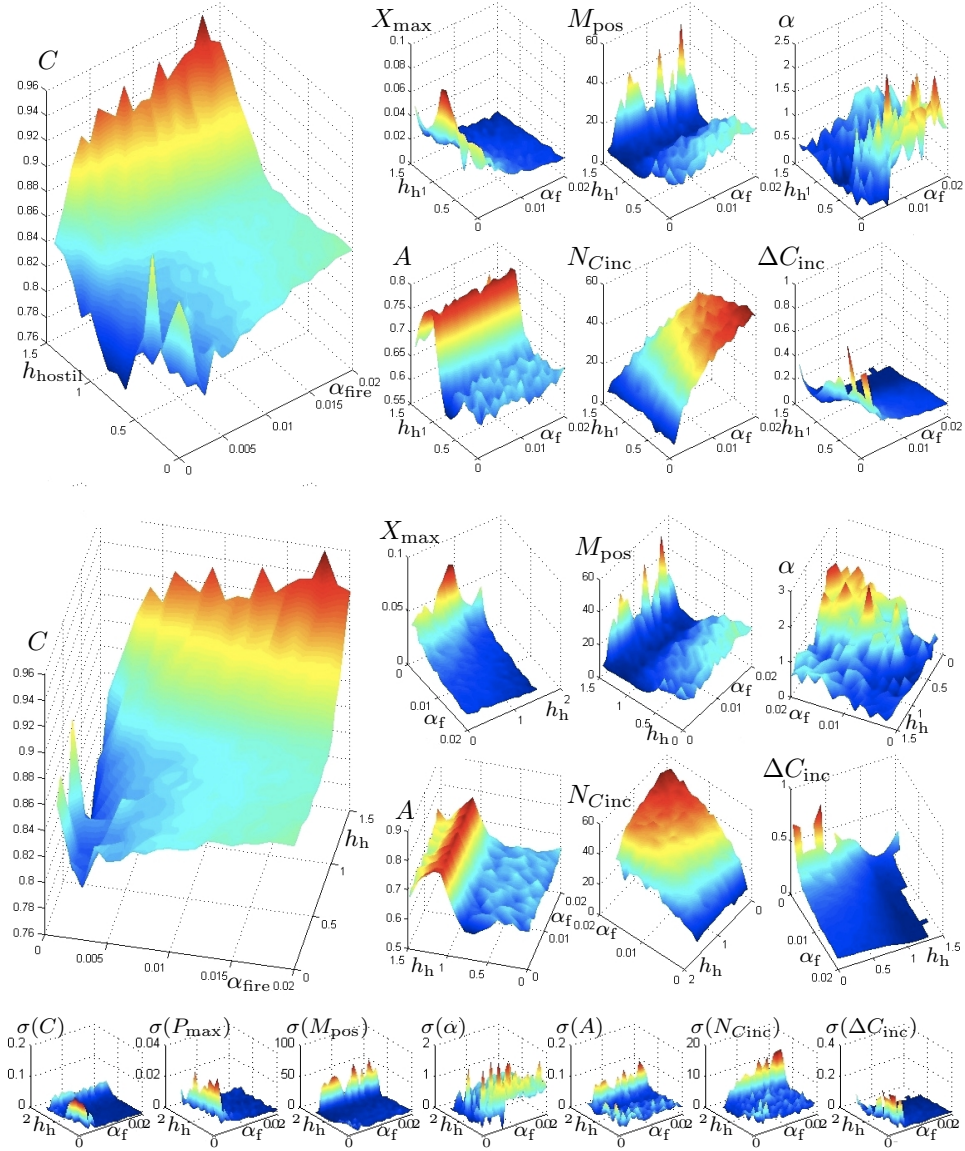
\begin{figure}
  \includegraphics[width=12.7cm]{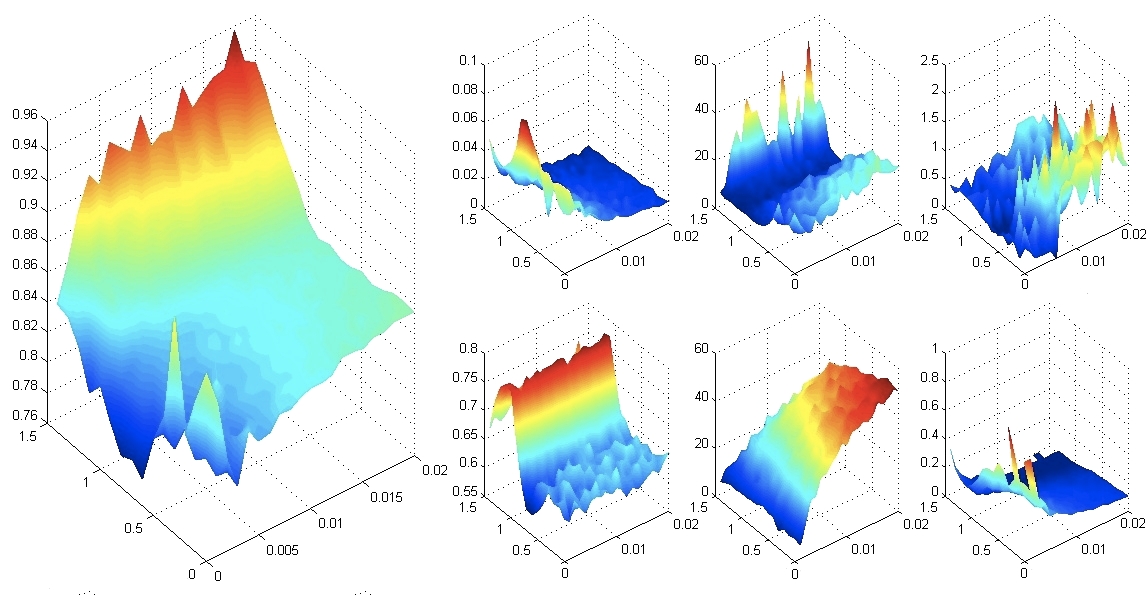}
  \put(-350,158){$C$}\put(-356,41){\small $h_{\rm hostil}$}\put(-240,35){\small $\alpha_{\rm fire}$}
  \put(-224,174){$X_{\rm max}$} \put(-150,174){$M_{\rm pos}$} \put(-70,174){$\alpha$}
  \put(-216,84){$A$} \put(-150,84){$N_{C\mathrm{inc}}$} \put(-78,84){$\Delta C_{\mathrm{inc}}$}
  %%%%%%%%%%%%%%%%%%%%%%%%%%%%%%%%%%%%%%%%%%
  \put(-214,111){\footnotesize $h_{\rm h}$}\put(-160,109){\footnotesize $\alpha_{\rm f}$}
  \put(-141,110){\footnotesize $h_{\rm h}$}\put(-88,109){\footnotesize $\alpha_{\rm f}$}
  \put(-68,110){\footnotesize $h_{\rm h}$}\put(-15,109){\footnotesize $\alpha_{\rm f}$}
  %%%%%%%%%%%%%%%%%%%%%%%%%%%%%%%%%%%%%%%%%%
  \put(-214,20){\footnotesize $h_{\rm h}$}\put(-160,18){\footnotesize $\alpha_{\rm f}$}
  \put(-141,20){\footnotesize $h_{\rm h}$}\put(-88,18){\footnotesize $\alpha_{\rm f}$}
  \put(-69,20){\footnotesize $h_{\rm h}$}\put(-16,18){\footnotesize $\alpha_{\rm f}$}\\[6pt]
  %%%%%%%%%%%%%%%%%%%%%%%%%%%%%%%%%%%%%%%%%%
  \includegraphics[width=12.7cm]{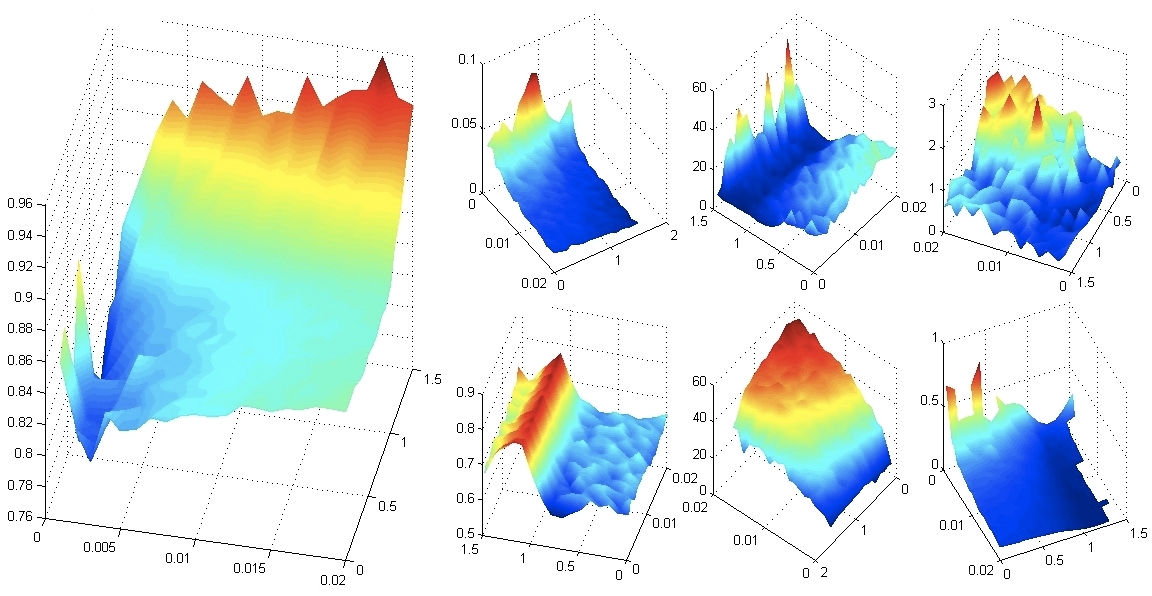}
  \put(-354,132){$C$}\put(-234,58){\small $h_{\rm h}$}\put(-278.5,4){\small $\alpha_{\rm fire}$}
  \put(-221,174){$X_{\rm max}$} \put(-147,168){$M_{\rm pos}$} \put(-70,161){$\alpha$}
  \put(-215,68){$A$} \put(-150,74){$N_{C\mathrm{inc}}$} \put(-76,84){$\Delta C_{\mathrm{inc}}$}
  %%%%%%%%%%%%%%%%%%%%%%%%%%%%%%%%%%%%%%%%%%
  \put(-158,107){\footnotesize $h_{\rm h}$}\put(-204,104){\footnotesize $\alpha_{\rm f}$}
  \put(-120,98){\footnotesize $h_{\rm h}$}\put(-88,116){\footnotesize $\alpha_{\rm f}$}
  \put(-22,102){\footnotesize $h_{\rm h}$}\put(-66,106){\footnotesize $\alpha_{\rm f}$}
  %%%%%%%%%%%%%%%%%%%%%%%%%%%%%%%%%%%%%%%%%%
  \put(-209,10){\footnotesize $h_{\rm h}$}\put(-156,29){\footnotesize $\alpha_{\rm f}$}
  \put(-100,10){\footnotesize $h_{\rm h}$}\put(-139,24){\footnotesize $\alpha_{\rm f}$}
  \put(-14,14){\footnotesize $h_{\rm h}$}\put(-62,14){\footnotesize $\alpha_{\rm f}$}\\[6pt]
  %%%%%%%%%%%%%%%%%%%%%%%%%%%%%%%%%%%%%%%%%%
  \includegraphics[width=12.7cm]{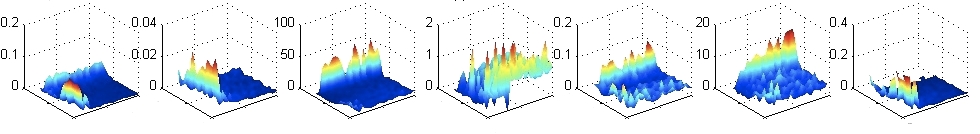}
  \put(-358,45){\footnotesize$\sigma(C)$}
  \put(-352,8){\footnotesize $h_{\rm h}$}\put(-324,6){\footnotesize $\alpha_{\rm f}$}
  \put(-356,11){\tiny$2$} \put(-336,2){\tiny$0$} \put(-316,9){\tiny$0\!.\!0\!2$}
  \put(-308,45){\footnotesize$\sigma(P_{\rm max})$}
  \put(-301,8){\footnotesize $h_{\rm h}$}\put(-273,6){\footnotesize $\alpha_{\rm f}$}
  \put(-305,11){\tiny$2$} \put(-285,2){\tiny$0$} \put(-265,9){\tiny$0\!.\!0\!2$}
  \put(-256,45){\footnotesize$\sigma(M_{\rm pos})$}
  \put(-250,8){\footnotesize $h_{\rm h}$}\put(-222,6){\footnotesize $\alpha_{\rm f}$}
  \put(-254,11){\tiny$2$} \put(-234,2){\tiny$0$} \put(-214,9){\tiny$0\!.\!0\!2$}
  \put(-205,45){\footnotesize$\sigma(\alpha)$}
  \put(-198,8){\footnotesize $h_{\rm h}$}\put(-170,6){\footnotesize $\alpha_{\rm f}$}
  \put(-202,11){\tiny$2$} \put(-182,2){\tiny$0$} \put(-162,9){\tiny$0\!.\!0\!2$}
  \put(-154,45){\footnotesize$\sigma(A)$}
  \put(-147,8){\footnotesize $h_{\rm h}$}\put(-119,6){\footnotesize $\alpha_{\rm f}$}
  \put(-151,11){\tiny$2$} \put(-131,2){\tiny$0$} \put(-111,9){\tiny$0\!.\!0\!2$}
  \put(-103,45){\footnotesize$\sigma(N_{C\mathrm{inc}})$}
  \put(-96,8){\footnotesize $h_{\rm h}$}\put(-68,6){\footnotesize $\alpha_{\rm f}$}
  \put(-100,11){\tiny$2$} \put(-80,2){\tiny$0$} \put(-60,9){\tiny$0\!.\!0\!2$}
  \put(-52,45){\footnotesize$\sigma(\Delta C_{\mathrm{inc}})$}
  \put(-46,8){\footnotesize $h_{\rm h}$}\put(-18,6){\footnotesize $\alpha_{\rm f}$}
  \put(-50,11){\tiny$2$} \put(-30,2){\tiny$0$} \put(-10,9){\tiny$0\!.\!0\!2$}
   \caption{ \small  Results of a sequence of simulation runs.
   Parameters are chosen as in figure~\ref{klfig:gridrun05}
   except for $\alpha_{\mathrm{Tvar}}=1$. Again the
   graphs are displayed for two different viewing positions. %\newline
   In this case, where temperature is independent of the
   carbon content $C$, one finds results not so different from
   those displayed in figure~\ref{klfig:gridrun05}. This, together
   with the significantly different results shown in
   figure~\ref{klfig:gridrun00} seems to indicate that
   the system characteristics change signifcantly for
   small values of $\alpha_{\mathrm{Tvar}}$, i.e. the mostly
   $C$-dominated case.}
  \label{klfig:gridrun10}
\end{figure}

To study these effects more closely, the authors have restricted the
grid to a smaller region in parameter space. Performing again several
simulation runs gave the results depicted in figure~\ref{klfig:gridrun05}.
In addition they changed the parameter $\alpha_{\mathrm{Tvar}}$ to zero
(figure~\ref{klfig:gridrun00}) and one (figure~\ref{klfig:gridrun10})
and find that the former case (where temperature is solely dominated
by $C$) is significantly different, while the latter case gives results very
similar to those obtained for $\alpha_{\mathrm{Tvar}}=\frac12$.

\clearpage

\subsection{Interplay of parameters}
\label{ssec:interplay}

The model parameters are not completely indepent.
For example it is clear that in figure~\ref{klfig:gridrunfull}
the population typically dies out for $h_{\mathrm{hostil}}\ge 1.75$.
One could expect, however, that a larger mutation rate $\mu$
may be beneficial in dealing with hostile environments and
that one could find values of $\mu$ that allow the population
to survive in most cases.

\begin{figure}
\begin{center}
  \includegraphics[width=12.5cm]{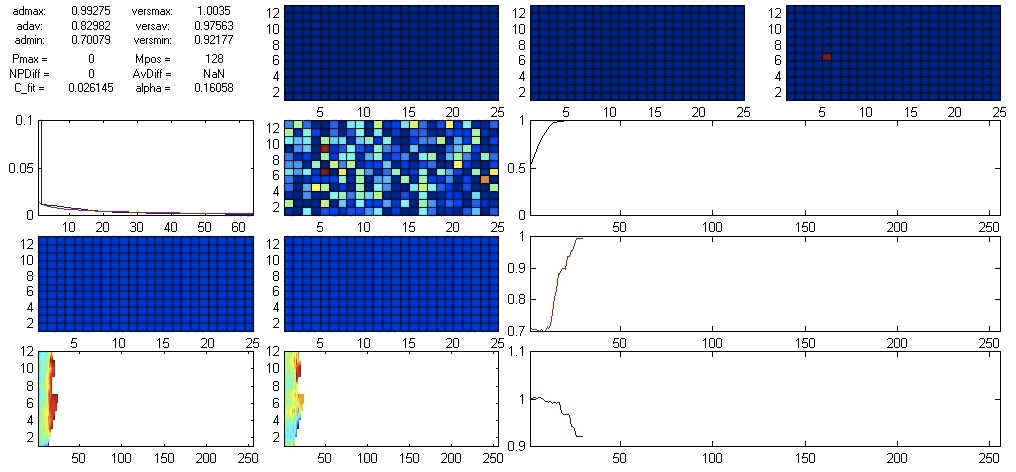} \\[3pt]
  \includegraphics[width=12.5cm]{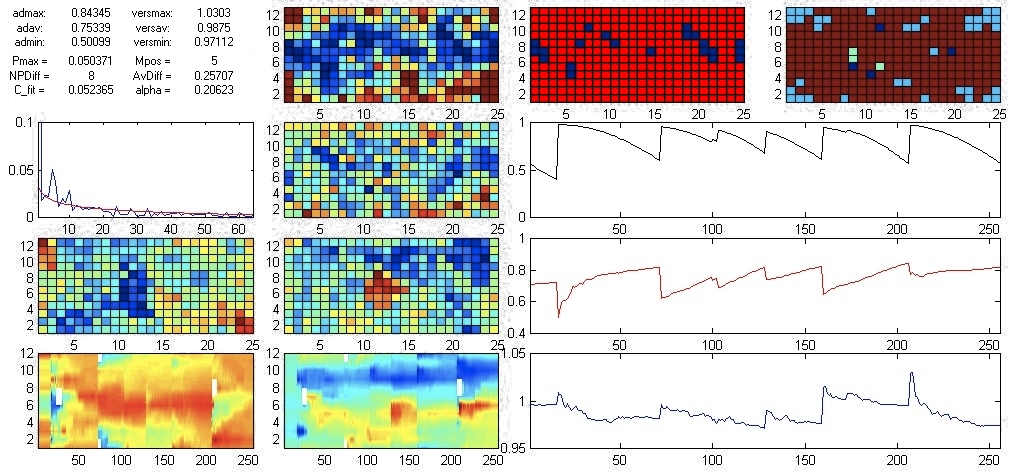} \\[3pt]
  \includegraphics[width=12.5cm]{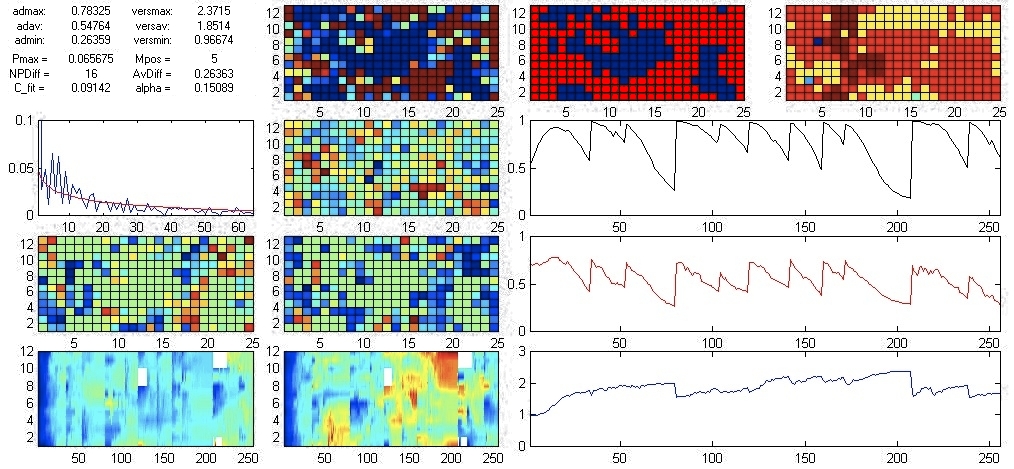}
\end{center}
  \caption{Large values for $h_{\mathrm{hostil}}$ can be
  partially compensated by increasing the mutation
  rate $\mu$. These plots show typical simulation runs
  for $h_{\mathrm{hostil}}=1.75$ and $\mu=0.1$, $\mu=0.3$
  and $\mu=0.5$. (Other parameters were chosen as in
  figure~\ref{klfig:simrun}.)}
  \label{klfig:mutation}
\end{figure}

This is indeed the case, as illustrated in figure~\ref{klfig:mutation}.
For $h_{\mathrm{hostil}}= 1.75$ and $\mu=0.3$ the population typically
survives. This is also true for $\mu=0.5$, though in this case it
repeatedly comes close to extinction. This is a hint that also in this
model a mutation rate which is too large can threaten the survival
of a species.

%%%%%%%%%%%%%%%%%%%%%%%%%%%%%%%%%%%%%
\section{Possible Extensions of the Model}
\label{klsec:carbclim_extension}

There are several possible extensions of this model, which
might be interesting to study:

\begin{itemize}
\item One could modify the grid by of introduction of uninhabitable
  spots, distributed randomly or as clusters of some fractal shape.
  One may also consider the growth of such uninhabitable spots
  (which can be interpreted as spreading of cities and deserts).

\item In a more complete model one could explicitely consider
  the influence of soil, which can store carbon from deceased plants.
  Another important factor is humidity, which can reduce the rate
  of forest fires and also has influence on temperature. This could
  help to make the model more realistic, but at the same time harder
  to analyze.
  
\item One could dismiss the normalization condition $C\le 1$ and
  increase the total amount of Carbon available in the system
  during the simulation run (to mimic the effect of burning
  fossile fuels).

\item One could try to explicitely include the effects of water vapor
  by setting up a a separate cycle for $\mathrm{H}_2\mathrm{O}$, coupled
  to the plant life grid, but endowed with a separate dynamics on
  a shorter timescale. It might even be worthwhile to think about
  coupling a Stochastic Cellular Automata Carbon climate model
  as proposed in this article to a more traditional model based on
  atmosphere and ocean dynamics.

\pagebreak

\item A fascinating extension may be the coupling of the system
  to an ``ocean'', i.e. a carbon sink with limited
  absorption rate and temperature-dependent capacity.
  
This may be especially important, since the solubility of $\mathrm{CO}_2$
in water decreases with increasing temperature\footnote{This is
in general true for the solubility of gases in liquids, while it is exactly
the other way round for the solubility of solids in liquids. This is easy
to understand, since enhanced thermal movement of the molecules
makes it easier for gases to evaporate, while it makes it harder for
solids to crystallize.}. So while the oceans can store enormous amounts
of $\mathrm{CO}_2$, their capacity decreases with global warming. This
could lead to a dramatic positive feedback chain:

In such a scenario, an increase of  temperatures above a certain critical
level could lead to a massive release of $\mathrm{CO}_2$ from the oceans
and thus a further rise of temperature. Confronted with this frightening
perspective, even simple models which allow to study the dynamics
of such catastrophic behaviour can be extremely valuable.
\end{itemize}

\noindent Apart from these extensions, there are also several aspects of the
original model that should be studied in greater depth. The interplay
of model parameters, as sketched in subsection~\ref{ssec:interplay},
certainly deserves closer examination. In addition it may be interesting
to study size-dependence, in particular the correspondence between
``critical'' $\alpha_{\mathrm{fire}}$ and the extension of the grid (for
example by considering the limit $N_iN_j\to \infty$, $\alpha_{\mathrm{fire}}\to 0$
with the product $N_iN_j\alpha_{\mathrm{fire}}$ held fixed).

%%%%%%%%%%%%%%%%%%%%%%%%%%%%%%%%%%%%%
\section{Summary and Conclusions}
\label{klsec:carbclim_summary}

The authors have studied a ``small-world'' model designed to describe
the interplay between global climate, carbon storage and sudden
outbreaks of forest fires.

They have seen that the least predictable behaviour is found for small, but
nonzero values for the probability $\alpha_{\mathrm{fire}}$ of a cell catching fire.
While in most cases an increased hostility of the environment leads to a
smaller average population and thus to a higher carbon content of the
atmosphere, this is not necessarily true in the most complex (and probably
most realistic) case of small, but nonzero $\alpha_{\mathrm{fire}}$.

This ``critical'' behaviour is more dramatic in the case of temperature
dominated by atmospheric carbon (instead of geographic parameters),
in terms of this model for $\alpha_{\mathrm{Tvar}}\approx 0$.

They have also seen that there exists an interesting interplay of parameters,
as for example the fact that a larger hostility of the environment can be
partially compensated by a higer mutation rate. In general, the model
exhibits a rich structure worth further investigation, and offers the possibility
of several interesting extensions.

\end{document}